\documentclass{ametsoc}
\journal{jcli}
\bibpunct{(}{)}{;}{a}{}{,}
\usepackage{caption}

\usepackage{amsmath,amsfonts,amssymb,bm}
\usepackage{mathptmx}%{times}
\usepackage{etoolbox,lineno}
\usepackage{graphicx}
\usepackage{natbib}
\usepackage{xcolor}
\title{Multiscale interactions between monsoon intra-seasonal oscillations and low pressure systems that produce heavy rainfall events of different spatial extents}

 \authors{Akshaya C Nikumbh\correspondingauthor{Akshaya C Nikumbh, nikumbh@iisc.ac.in}$^{1,2}$, Arindam Chakraborty$^{1,2}$, G.S. Bhat$^{1,2}$, Dargan M. W. Frierson $^{3}$}

\affiliation{\nolinenumbers $^{1}$Centre for Atmospheric and Oceanic Sciences, Indian Institute of Science, Bangalore, 560012, India.\\
$^{2}$Divecha Centre for Climate Change, Indian Institute of Science, Bangalore, 560012, India.\\
$^{3}$Dept. of Atmospheric Sciences, University of Washington, Seattle, 98195, USA.}

\email{nikumbh@iisc.ac.in}

\abstract{ \nolinenumbers The sub-seasonal and synoptic-scale variability of the Indian summer monsoon rainfall are controlled primarily by monsoon intra-seasonal oscillations (MISO) and low pressure systems (LPS), respectively. The positive and negative phases of MISO lead to alternate epochs of above-normal (active) and below-normal (break) spells of rainfall. LPSs are embedded within the different phases of MISO and are known to produce heavy precipitation events over central India. Whether the interaction with the MISO phases modulates the precipitation response of LPSs, and thereby the characteristics of extreme rainfall events (EREs) remains unaddressed in the available literature. In this study, we analyze the LPSs that produce EREs of various spatial extents viz., Small, Medium, and Large over central India from 1979 to 2012. We also compare them with the LPSs that pass through central India and do not give any ERE (LPS-noex).  We find that thermodynamic characteristics of LPSs that
trigger different spatial extents of EREs are similar. However, they show differences in their dynamic characteristics. The ERE producing LPSs are slower, moister and more intense than LPS-noex.  The LPSs that lead to Medium and Large EREs tend to occur during the positive phase of MISO when an active monsoon trough is present over central India. On the other hand, LPS-noex and the LPSs that trigger Small EREs occur mainly during the neutral or negative phases of the MISO. The  large-scale dynamic forcing, intensification of LPSs, and diabatic generation of low-level potential vorticity due to the presence of active monsoon trough help in the organization of convection and lead to Medium and Large EREs. On the other hand, the LPSs that form during the negative or neutral phases of MISO do not intensify much during their lifetime and trigger scattered convection, leading to EREs of small size. }%These results connote a modulation of  the precipitation response of monsoon lows by intra-seasonal oscillations.}

\begin{document}
\nolinenumbers
%% Necessary!
\maketitle

 \section{Introduction}
 Though there are significant advancements in the understanding of heavy rainfall events over the past couple of decades, predicting their location and intensity with high accuracy and sufficient lead time has remained a challenging task. The challenges come mainly from identifying favourable initial conditions,  feedbacks, external drivers, and nonlinear processes that lead to the development of extreme events \citep{sillmann2017understanding}. The interaction of several spatio-temporal scales during the Indian summer monsoon \citep{ding2007variability, singh2019indian} poses unique challenges to decipher these processes.
 
 During the Indian summer monsoon, the sub-seasonal variability comes mainly from the monsoon intra-seasonal oscillations (MISO) \citep{yasunari1979cloudiness, sikka1980maximum,goswami2001intraseasonal,webster1998monsoons,wang2006asian}. The MISO consist of two dominant modes, 10 to 20 day \citep{krishnamurti1976oscillations} and 30 to 60 day oscillations \citep{yasunari1979cloudiness, sikka1980maximum} that account for 25\% and 66\% of total intra-seasonal variability, respectively \citep{annamalai2001active}. The 30 to 60 day mode, which originates over the Equatorial Indian ocean travels northward to the Indian landmass, whereas the 10 to 20 day mode originates over the South China Sea and propagates north-westward \citep{webster1998monsoons,wang2006asian}. The different phases of these two modes and their interactions lead to alternate periods of above (active) and below (break) normal rainfall over India. Conditions conducive for heavy rainfall events are formed when both modes of the MISO are in phase and exhibit simultaneous positive anomalies \citep{karmakar2015decreasing}. \citet{francis2006intense} showed that heavy precipitation events over the Western Ghats are associated with the tropical convergence zone that forms during an active phase of the MISO. \citet{karmakar2015decreasing} showed that the maximum number of extreme rainfall events (EREs) over central India occur during an active phase of low-frequency MISO (30 to 60 day mode). With the changing global climate, these intra-seasonal oscillations of the monsoon are expected to become more extreme and increase the likelihood of EREs over India \citep{turner2009subseasonal}. The signs of these changes are already becoming evident over the past couple of decades. The active spells are becoming intense and the dry spells more frequent \citep{singh2014observed}. The influence of the MISO  on heavy rainfall events over India has been speculated  before \citep{turner2009subseasonal, francis2006intense, singh2014observed,karmakar2015decreasing}, however, its exact impact on the characteristics of heavy rainfall events has not been elucidated in detail.
 
 Embedded within the intra-seasonal oscillations are synoptic disturbances called monsoon low pressure systems (LPSs). LPSs form mainly over the Bay of Bengal and  travel north-westward, bringing the episodes of heavy rainfall over central India \citep{godbole1977composite, Sikka1977, Patwardhan2020}. \citet{ajayamohan2010increasing} defined a normalized index called the synoptic activity index using the intensity, frequency, and duration of LPSs. They showed that the number of EREs over central India and the synoptic activity index show a strong correlation, and suggested that the increasing synoptic activity possibly explains the rising trend of EREs. Similarly, \citet{karmakar2015decreasing} found that the relative contribution from the synoptic variability (3 to 9 days) to the variance of daily rainfall over central India has been increasing. \citet{nikumbh2020large} showed that two-third of  LPSs that travel through central India give at least one ERE. They also showed that the EREs with the large spatial extent ($area \ge 70,000 km^{2}$) are initiated by the interaction between an LPS and a secondary cyclonic vortex (SCV). When an SCV is located to the west of an LPS, the conditions favourable for large-scale EREs are formed in between these two cyclonic vortices. Studies attribute either increases in the number of LPSs days \citep{pai2016active} or increases in the number of monsoon lows \citep{krishnamurthy2011extreme} to increases in the frequency of EREs over central India. It has been shown that the different phases of MISO influence the formation and propagation of LPSs \citep{goswami2003clustering,sikka2006study}. \citet{goswami2003clustering} observed the clustering of LPSs during an active phase of the MISO. They showed that LPSs are 3.5 times more likely to form during the active phase of the MISO than the break phase. Also, the tracks of LPSs during the active phase congregate mainly along the monsoon trough that is situated over central India. On the other hand, the LPSs that form during the break phase are confined to the Himalayan foothills. Does the interaction of the MISO and LPSs also play a role during heavy rainfall events? The available literature does not address this question.
 
 \citet{nikumbh2019recent} showed that the EREs over central India exhibit different spatial scales and classified them as Small, Medium and Large EREs. They showed that the daily EREs of all sizes show a strong association with LPSs and occur mainly to the west of their centre. They also noted that the EREs of different sizes have different large-scale conditions and speculated that the underlying physical processes could be different for various sizes of EREs. Using a k-mean clustering on an hourly rainfall data, \citet{moronstorm} classified wet events over India into different  storm types and grouped them into two broad categories. The first group with short space-time scales and intense rainfall is phase-locked with the diurnal cycle i.e., triggered mostly in mid-afternoon. The other group, with a longer duration, larger spatial extent, and less intense rainfall does not show any peak triggering time, so is possibly associated with large mesoscale systems or tropical lows. These studies show that the governing processes of EREs may differ based on the spatial extent of EREs and it requires further investigation.
 
 In this study, our main objective is to understand the characteristics and background conditions that determine the response of  LPSs while producing EREs over central India with different spatial extents viz., Small, Medium, and Large. We speculate that the different phases of the MISO could modulate the precipitation response of LPSs. So, we explore the possible influence of the MISO phases on the spatial extent of EREs produced by LPSs. 
 
% \subsection*{subsection}
% text...
 \section{Method}
  EREs of the summer monsoon months (June to September) for the period of 1979-2012 are identified using the daily gridded rainfall dataset developed by the India Meterological Department \citep{rajeevan2006high}. This dataset has a spatial resoultion of $1^{\circ}\times1^{\circ}$. Central India ($15^{\circ}$-$25^{\circ}$N, $75^{\circ}$-$85^{\circ}$E inset box in \textcolor{black}{Fig. 1a}) is our study area.
 
 Following \citet{nikumbh2019recent}, an ERE at each grid is identified using a $99.5^{th}$  threshold and then using the connected component labelling analysis \citep{falcao2004image}, the neighbouring grids with extreme rainfall are combined. This gives the EREs of different sizes, where the size is measured in  $1^{\circ}\times1^{\circ}$  grid units. The EREs are then classified according to their spatial extents as Small ($size = 1$, area $\lesssim$ $10^{4}$ $km^{2}$), Medium ($size$ $2-5$, $10^{4}$ $km^{2}$ $\gtrsim$ area $\lesssim$ $7 \times 10^{4}$ $km^{2}$), and Large ($size \ge 6 $, area  $\gtrsim$  $7 \times 10^{4}$ $km^{2}$) EREs. Our classification of EREs into three types is based on the results reported in \cite{nikumbh2019recent} and \cite{nikumbh2020large}. It mainly considers the frequency of occurrence and synoptic signatures into account.\\ 
 To avoid a possible repeat count of the same event, the EREs are checked iteratively for two consecutive days. If any two EREs occur on consecutive days, the larger size ERE is considered and the smaller size ERE is discarded. If an ERE of the same category occurs on the next day, then only an ERE of the first day is considered. Over central India, the probability of getting Small EREs is the highest (Probability=71.4\%, \textcolor{black}{Fig. 1a}) and this probability decreases as the size of ERE increases. The average probability of occurrences of  Medium and Large EREs are 26.2\% and 2.4\%, respectively. The Medium and Large events are not only bigger in the spatial extent but also slightly more intense (\textcolor{black}{Fig. 1b}). The average rainfall at the centre of Small, Medium, and Large ERE is 122 $mm$ $day^{-1}$, 135 $mm$ $day^{-1}$ and 145 $mm$ $day^{-1}$, respectively. 
 
 The dataset by Hurley and Boos \citep{hurley2015global}, which is available from 1979 to 2012 is used to identify LPSs. The presence of an LPS is checked for a given day of an ERE over the domain of $75^{\circ}$E-$90^{\circ}$E, $15^{\circ}$N-$30^{\circ}$N. As shown in \textcolor{black}{Fig. 1a}, the EREs of central India show a strong association with LPSs. The EREs of all sizes over central India have an LPS present for more than 80\% of the time. This association is stronger for the bigger size EREs where all Large EREs are associated with LPSs. If there are multiple LPSs on the same day, only the nearest LPS to an ERE is considered for the analysis. In order to address our objective, we consider only those EREs that had an LPS present. We also examined the LPSs that did not produce any ERE over central India to bring out the contrast. The LPSs which lead to Small, Medium, and Large EREs are referred to as LPS-sm (number of cases(n)=118), LPS-med (n= 133) and LPS-lg (n=20), respectively (abbreviations listed in \textcolor{black}{Table 1}). The LPSs that pass through the study region and did not lead any ERE are named as LPS-noex (n=166). 
 
 The  meteorological conditions are studied using the ERA-Interim reanalysis \citep{dee2011era}. The lag composites are calculated with respect to the day of an ERE (Day-0) for ERE producing LPSs. For LPS-noex, Day-0 indicates the genesis day of an LPS. The daily anomalies  of meteorological fields are obtained by subtracting the daily climatology. The significance of anomalies is checked using the two-sided t-test. The cumulative distribution functions (CDFs) and boxplots are compared using the Kolmogorov-Smirnov test (K-S test). The 95\% confidence intervals for the line plots are calculated by generating $10^{4}$ bootstrapping samples.

%vs

\section{Results} %\label{sec:results}
 \subsection{Characteristics} %\label{sub:chars}
 The characteristics of LPSs such as relative vorticity, speed of propagation, and total column water vapor are evaluated along their entire path at a 6-hourly time step (\textcolor{black}{Fig 2}). The relative vorticity ($\zeta$ at 850 hPa)  is lowest for LPS-noex ($average=4 \times 10^{-5}$ $s^{-1}$) and gradually increases from LPS-sm ($average=5 \times 10^{-5}$ $s^{-1}$)  to LPS-med ($average=6 \times 10^{-5}$ $s^{-1}$), and finally is the highest for LPS-lg ($average=8 \times 10^{-5}$ $s^{-1}$).  The LPSs that produce EREs have a significantly slower speed of propagation than LPS-noex ($pvalue < 0.01$). Although the LPS-sm tend to be faster, the speed of propagation of ERE producing LPSs is not much different from each other ($pvalue > 0.05$). The average speed of ERE producing LPSs is 4.6 $m$ $s^{-1}$, whereas LPS-noex have an average speed of 5.3 $m$ $s^{-1}$. LPS-noex are significantly drier than ERE producing LPSs and they get progressively wetter from LPS-sm to LPS-lg ($pvalue$ for each pair of LPS $< 0.01$). The average total column water vapor of LPS-noex, LPS-sm, LPS-med and LPS-lg is 51, 61,63 and 65 ($kg$ $m^{-2}$), respectively.
 
 Further, we look at the distribution of pressure velocity ($\omega$ at 500 hPa), relative humidity ($RH$), and equivalent potential temperature ($\theta_{e}$) (\textcolor{black}{Fig. 3}). The absolute value of $\omega$ increases progressively from Small to Medium and is maximum for Large EREs (\textcolor{black}{Fig. 3a}). The location of an ERE is marked by the region of maximum $\omega$, and its spread determines the spatial extent of the ERE. In the coordinate system of an LPS, the EREs are located to the western side of the centre of an LPS \citep{nikumbh2019recent}. Thus, the upward motion in the midtroposphere on the eastern side of EREs indicate the updraft associated with LPSs. For Large EREs, the ascending zone is present even to the west of the ERE due to the presence of an SCV \citep{nikumbh2020large}.
 
 At low-levels, the atmosphere is mostly saturated for all types of EREs (\textcolor{black}{Fig. 3b}). Below 700 hPa  to the west of EREs, the low level jet (LLJ) provides the moisture that saturates the low-level atmosphere. At upper levels, the air that comes from the dry northwest landmass of India is less humid. On the other hand, the air on the east side, which comes with LPSs from the Bay is more humid even at upper levels. The interesting difference among different EREs, is the presence of a deep saturated air layer that extends up to the mid-troposphere near the centre of Medium and Large EREs. The deep saturated layer helps to sustain the convection by providing higher plume buoyancies~\citep{bretherton2004relationships,holloway2009moisture}. This deep humid layer is absent for Small EREs. The spread of the saturated layer at low levels does not seem to be playing a role in determining the spatial extent of EREs. Similar to $RH$,  the distribution of $\theta_{e}$ at 950-hPa does not determine the extent of EREs (\textcolor{black}{Supplementary Fig. 1}). The median value of $\theta_{e}$  and its distribution for all EREs is almost similar (\textcolor{black}{Fig. 3c}, $pvalue$ using K--S test $>0.05$). On the other hand, the distribution of $\omega$ at 500 hPa is significantly different for all (\textcolor{black}{Fig. 3d}, $pvalue$ using K--S test$<0.05$) and its spread matches with the spatial extent of EREs (\textcolor{black}{Fig. 3a}). 
 
 The changes  in precipitation  are often investigated in terms of dynamic and thermodynamic characteristics \citep{trenberth2003changing,allan2008atmospheric,emori2005dynamic}. The circulation related characteristics are known as dynamic (eg., $\omega$, $u$, $v$) and those related to temperature or moisture are called thermodynamic characteristics (eg., $RH$, static stability). From the primary analysis, we see that the dynamic characteristics play a key role in determining the spatial extent of EREs, whereas the spread of thermodynamic quantities does not seem to be determining the location and spatial extent of EREs.

 \subsection{Background conditions} %\label{sub:back}
 The lagged composites of geopotential height bring out the distinct background conditions for LPSs that lead to the EREs of different spatial extents (\textcolor{black}{Fig. 4 a-c}). Zonally elongated anomalies are observed from two days before the occurrence of Large and Medium EREs, indicating the active monsoon conditions. They are not seen prior to Small EREs and during the genesis of LPS-noex (\textcolor{black}{Supplementary Figure 2a}).  For Large EREs, there is also a signature of an SCV over the west coast. During the break, the rainfall over central India is subdued but it is enhanced over the south eastern India \citep{ramamurthy1969monsoon, pai2016active}. It exhibits the anomalous low over the south eastern peninsula similar to the observed anomalies for LPS-sm (\textcolor{black}{Fig. 4 a}) and LPS-noex  (\textcolor{black}{Supplementary Fig. 2a}). 
 
 The presence of an active monsoon conditions can be confirmed further by the rainfall over the core monsoon zone \citep{rajeevan2010active}. The rainfall anomalies are positive over the core monsoon zone before the occurrence of Large and Medium EREs (\textcolor{black}{Fig. 5}), indicating an active monsoon spell. The active spell during Large EREs is more intense than that of Medium EREs. For Small EREs and LPS-noex the rainfall over central India is below or close to normal. On the day of ERE, the rainfall anomalies peak over central India and then slowly start falling. As Day(0) for LPS-noex is the day of genesis, the peak in the rainfall over central India comes slightly later (Day(+3)), when the LPS makes landfall and travels over central India (\textcolor{black}{Supplementary Fig. 2b}).
 
 To confirm an association of EREs with the phase of the MISO, we define the MISO index following \citet{goswami2003clustering}. The relative vorticity at 850 hPa is averaged over the core monsoon zone ($75^{\circ}-90^{\circ}$E,$15^{\circ}-25^{\circ}$N) from May to October of each year from 1979 to 2012. These anomalies are then detrended first and filtered for 10 to 90 days. To avoid the edge effects we use the Tukey window. The filtered values are then normalized by the standard deviation of the relative vorticity for the entire time period to get the MISO index. The kernel density estimate of the MISO index is shown in \textcolor{black}{Fig. 6}. The LPS-lg and LPS-med show a peak at the positive MISO index while LPS-sm peaks at the negative MISO index. The 70\% and 65\% of LPS-lg and LPS-med occur during the positive phase of the MISO. On the other hand, 80\% and 84\% of LPS-sm and LPS-noex occur during the neutral to negative phase of the MISO (MISO index $\le 0.5$) . This consolidates our speculation of a positive MISO phase for LPS-lg and LPS-med, and a neutral or negative phase for LPS-sm and LPS-noex.
 
 The preferential location where the EREs of different sizes occur follows the location of the monsoon trough during the active and break phases of the MISO. The trough shifts northward near the foothills during the break phase and it is situated over the core monsoon zone during an active phase  \citep{rao1976southwest, ramamurthy1969monsoon}. Small EREs tend to occur near the northern boundary of the study region (North of $22^{\circ}$N), while Medium to Large EREs occur mainly in the southern part ($15^{\circ}$-$20^{\circ}$N) (Fig. 2b and supplementary Fig. 5 of \citet{nikumbh2019recent}). This corroborates again the influence of the MISO on the different sizes of EREs. 
 
 The background conditions reveal that Large and Medium EREs occur mainly during the positive (active) phase of the MISO, while Small EREs tend to occur during the neutral or negative phase of the MISO. By looking at the forcing responsible for the distribution of $\omega$ and the organization of convection, we can understand the processes that control the spatial extent of  EREs. In the next sections, we analyze these processes by looking at dynamic forcing, vortex intensification, and diabatic generation of potential vorticity.

\subsection{Dynamic forcing}% \label{sub:dynf}
By using the quasi geostrophic (QG) omega equation \citep{hoskins1978new} we examine the dynamic forcing that leads to the observed differences in the distribution of vertical velocity. We use the modified definition of Q-vector ($-2 \nabla \cdot Q$) \citep{kiladis2006three,phadtare2019characteristics, boos2015adiabatic} for the tropics, where the geostrophic wind is replaced by the rotational wind. This modified form of $\nabla \cdot Q$ is given by,
\begin{equation}
\frac{p}{R}\nabla \cdot Q \simeq \frac{\partial T}{\partial x} \frac{\partial \zeta}{\partial y}- \frac{\partial T}{\partial y} \frac{\partial \zeta}{\partial x}-\frac{\partial^2 T}{\partial x \partial y} \left (\frac{\partial^2 \psi}{\partial^2 x} -\frac{\partial^2 \psi}{\partial^2 y} \right)+\frac{\partial^2 \psi}{\partial x \partial y} \left (\frac{\partial^2 T}{\partial^2 x} -\frac{\partial^2 T}{\partial^2 y} \right)  
\end{equation} 

where $p$, $R$, $T$, $\zeta$ , $\psi$   are pressure, the gas constant for dry air, temperature, the vertical component of relative vorticity and the stream-function of the horizontal wind, respectively and other terms have their usual meaning. \textcolor{black}{Fig. 7} shows the evolution of dynamic forcing ( $-2 \nabla \cdot Q$  at 500 hPa) over central India before and after the occurrence of EREs of different sizes. The large increase in the dynamic forcing over the entire central India is observed from two days before Medium and Large EREs. The enhanced dynamic forcing over central India for Medium and Large EREs is due to the presence of an active monsoon conditions (\textcolor{black}{Fig. 4}). During an active monsoon, the enhanced relative vorticity over the core monsoon zone and the negative vorticity anomalies over the foothills exist, which are also observed for LPS-med and LPS-lg (\textcolor{black}{Supplementary Fig. 3}). The presence of a zonally elongated monsoon trough enhances the vorticity and temperature (\textcolor{black}{Supplementary Fig.3}) that sets up large-scale dynamic forcing over central India for Medium and Large EREs. For LPS-sm and LPS-noex, the large-scale dynamic forcing is absent. For Small EREs, the anticyclonic vorticity and cold anomalies observed over central India, which creates weak dynamic subsidence over central India before the arrival of LPSs. When an LPS-sm enters central India large-scale dynamic forcing is absent, and it barely overcomes the background dynamic subsidence. The LPS-noex register a slight increase in the dynamic forcing over central India, 3-days after its genesis, coincident with its landfall over central India. 

\subsection{Vortex intensification} %\label{sub:vortb}
The LPS-lg and LPS-med are not only stronger compared to LPS-sm and LPS-noex during their genesis but also continue to intensify throughout their lifetime (\textcolor{black}{Fig. 8a}). The LPS-lg has the highest growth rate followed by LPS-med. The LPS-sm and LPS-noex have very small growth rates. The vortex stretching is the major source of vorticity at low-levels for LPSs  \citep{ daggupaty1977vorticity,godbole1978cumulus,rajamani1989some}. It is especially high for LPS-lg and LPS-med (\textcolor{black}{Fig. 8b}). During the positive phase of MISO, the strength of the LLJ is high and a strong wind exists over the southern boundary of the monsoon trough (\textcolor{black}{Supplementary Fig. 4a}). The interaction of the LLJ with LPS-lg and LPS-med results in the formation of  the strong convergence zone to the west of LPSs (\textcolor{black}{Supplementary Fig. 4b}), which results in the vortex stretching and its intensification. This shows that the positive phase of MISO helps in the intesification of the LPS-lg and LPS-med. On the other hand, LPS-sm and LPS-noex do not intensify as much because the LLJ is weak during the negative and neutral phases of MISO.

\subsection{Diabatic generation of potential vorticity} % \label{sub:pv}
We look at the  influence of  an active monsoon trough  in strengthening the LPS-lg and LPS-med through the diabatic effects by examining potential vorticity (PV). An approximate diabatic generation of PV  is given by \citep{martin2013mid},
\begin{equation}  \label{eq:3}
\frac{d}{dt} (PV) \approx  -g (\zeta+f) \frac{\partial}{\partial p} \left(\frac{d\dot{Q}}{dt}\right)
\end{equation}

where terms  $PV$, $g$, $\zeta$, $f$, and $\frac{d\dot{Q}}{dt}$ are potential vorticity, gravitational constant, relative vorticity, Coriolis parameter and diabatic heating, respectively. The $PV$ is generated where the vertical gradient of diabatic heating term is positive and destroyed where it is negative. We calculate the diabatic heating term ($\frac{d\dot{Q}}{dt}$), which is a residual term of the thermodynamics equation \citep{holton2012introductiondia, ling2013diabatic} as follows: 
\begin{equation}  \label{eq:4}
\left(\frac{d\dot{Q}}{dt}\right) = \frac{T}{\theta} \left( \frac{\partial \theta}{\partial t} + u \frac{\partial \theta}{\partial x}+ v \frac{\partial \theta}{\partial y}+ \omega \frac{\partial \theta}{\partial p}  \right)
\end{equation}

The vertical profile of diabatic heating averaged over central India is shown from Day(-15) to Day(0) in \textcolor{black}{Fig. 9}. About 15 days before the EREs, the vertical profile of diabatic heating is almost same for all cases. As noted in the earlier studies, the mid-level warming is generated by convection \citep{ajayamohan2016role, hazra2015space,yanai1973determination}. The low level and upper-level cooling are associated with evaporative cooling and radiative cooling, respectively. The distinction among the profiles starts from Day(-9) when the rainfall anomalies (\textcolor{black}{Fig. 5}) over central India for Large and Medium EREs slowly start picking up. The mid-tropospheric diabatic heating continues to intensify for Large and Medium EREs in tandem with the rainfall anomalies over central India. The maximum anomalies in diabatic heating at low levels are observed a day before the Large EREs. For Medium EREs, the anomalies of diabatic heating are almost similar on the day of an ERE and a day before. For Small EREs, they peak on the event day. The vertical gradient in the diabatic heating leads to a positive PV tendency in the lower troposphere and negative tendency in the upper troposphere (\textcolor{black}{Fig. 10}). A day before Large and Medium EREs, the diabatic PV generation deepens and reaches up to the mid-troposphere. Hence, the cyclonic circulation at low to mid-levels is strengthened over central India partly due to diabatic heating during the positive MISO phase for LPS-lg and LPS-med. This strengthened circulation can combine the scattered convection, and form a large-scale organized convection, leading to a large spatial extent of EREs.

 \section{Summary and Discussion}
 
 In this study, we investigate the characteristics and background conditions of the monsoon lows (LPSs) that lead to extreme rainfall events (EREs) of different spatial extents viz., Small ($size=1$, LPS-sm), Medium ($size=2-5$, LPS-med), and Large ($size\ge6$, LPS-lg) EREs over central India. We also compare them with the LPSs that do not lead to any ERE (LPS-noex). 
 
 The characteristics (Section 3a) of these LPSs reveal that the ERE producing LPSs are slower, moister, and more intense than the LPSs that do not give EREs (\textcolor{black}{Fig. 2}). The spatial extent and the location of EREs is coincident with the mid-tropospheric $\omega$ maximum. The spread of thermodynamic entities such as relative humidity or equivalent potential temperature does not seem to be controlling the spatial extent of EREs. This possibly indicates that the spatial extent of EREs is determined primarily by the dynamic characteristics (\textcolor{black}{Fig. 3}). 
 
 The analysis of background conditions (Section 3b) shows that the LPSs that set off Medium and Large EREs occur mainly during the positive phase of the MISO (\textcolor{black}{Fig. 4-6}), which forms an active monsoon trough over central India. On the other hand, the LPSs that form during the negative and neutral phases of the MISO do not have an active monsoon trough over central India and produce Small EREs. The LPSs that do not lead to any ERE also occur mainly during the neutral to negative phase of the MISO. The spatial distribution of the EREs of various sizes follow the spatial variations in the monsoon trough during the different phases of the MISO. The trough is situated over central India during the positive phase of the MISO and shifts near the foothills during the negative MISO phase. Accordingly, Large and Medium EREs occur mainly over the southern boundary ($15^{\circ}$N-$20^{\circ}$N) of central India, while Small EREs tend to occur near the northern boundary (north of $22^{\circ}$N).
 
 Processes (Section 3c-e, \textcolor{black}{Fig. 7-10})  through which the phases of the MISO modulate the response of LPSs in producing EREs of different sizes are as follows.  \\	
 1. During the positive phase of MISO, an active monsoon trough and a strong LLJ exists over central India. A large-scale dynamic forcing is set up due to enhanced temperature and vorticity gradients. When an LPS passes through central India it intensifies by vortex stretching as a strong convergence zone forms to the west of the LPS resulting from the interaction of the LLJ with the LPS. An active monsoon trough generates potential vorticity at low to mid-troposphere through the diabatic effects and strengthens the low-level cyclonic circulation. These conditions help in the organization of convection and LPSs produce medium and large EREs.\\
 2. During the negative phase of the MISO, large-scale dynamic subsidence exists at mid-levels over central India due to cold and negative vorticity anomalies. These LPSs do not intensify much in their lifetime. In the absence of an active monsoon trough, the low-level circulation is not strengthened via diabatic generation of potential vorticity. When an LPS passes through central India in such conditions, it triggers localized heavy rainfall events (Small EREs).\\
 %3. The LPSs that do not produce extreme rainfall events (LPS-noex)  are significantly weaker, faster, and drier than the ERE-producing LPSs. LPS-noex occur mainly during the neutral to negative phases of the MISO. Similar to LPS-sm, LPS-noex do not intensify and the low-level circulation over central India is not strengthened  in the absence of an active monsoon trough. 
 
Here, we show how the interaction of the MISO phases with LPSs can modulate the spatial characteristics of heavy rainfall events. Predicting in/out of phase conditions of multiple monsoon components can improve the forecast of different spatio-temporal characteristics of heavy rainfall events. Such forecast that includes the spatio-temporal characteristics could prove useful in preparing a proper action plan.
 
We use LPS-noex to bring out the contrast, however, further investigation is needed to understand them thoroughly and examine for the different spatio-temporal characteristics of wet events. A comprehensive study of the multi-scale interactions of different monsoon components for all wet and dry events is essential. Similarly, temporal characteristics of wet and dry events in addition to spatial characteristics could be explored with a high resolution dataset.  
 
 The monsoon lows and MISO are modulated by other tropical variability like the Madden Julian oscillations (MJO) \citep{haertel2017global, pai2011impact}, and the slow-varying (interannual) modes like ENSO and IOD \citep{ajayamohan2008influence, yoo2010analysis}. Further efforts are needed to understand the influences of these large-scale drivers on the characteristics of heavy rainfall events and thereby help to improve their prediction.  

%%%%%%%%%%%%%%%%%%%%%%%%%%%%%%%%%%%%%%%%%%%%%%%%%%%%%%%%%%%%%%%%%%%%%
% ACKNOWLEDGMENTS
%%%%%%%%%%%%%%%%%%%%%%%%%%%%%%%%%%%%%%%%%%%%%%%%%%%%%%%%%%%%%%%%%%%%%
\acknowledgments
A. C. and G. S. B. acknowledge funding from the DST, MoES and MoEF, Government of India. D. M. W. F. is
supported by NSF Grant AGS-1665247. We thank S. Paleri and C. Jalihal for their inputs.
%%%%%%%%%%%%%%%%%%%%%%%%%%%%%%%%%%%%%%%%%%%%%%%%%%%%%%%%%%%%%%%%%%%%%
% DATA AVAILABILITY STATEMENT
%%%%%%%%%%%%%%%%%%%%%%%%%%%%%%%%%%%%%%%%%%%%%%%%%%%%%%%%%%%%%%%%%%%%%
% 
%
%\datastatement
%The rainfall data used in the study is obtained from the IMD (\url{http://imdpune.gov.in/Clim_Pred_LRF_New/Grided_Data_Download.html}). The global monsoon LPSs track data set is available at \url{https://worldmonsoons.org/global-monsoon-disturbance-track-dataset/}. The ERA-interim data set is available online (\url{https://apps.ecmwf.int/datasets/data/interim-full-daily/}).
%\bibliography{references}
%%%%%%%%%%%%%%%%%%%%%%%%%%%%%%%%%%%%%%%%%%%%%%%%%%%%%%%%%%%%%%%%%%%%%
% REFERENCES
%%%%%%%%%%%%%%%%%%%%%%%%%%%%%%%%%%%%%%%%%%%%%%%%%%%%%%%%%%%%%%%%%%%%%
% Make your BibTeX bibliography by using these commands:
 %\bibliographystyle{ametsoc2014}
 \bibliography{references}
 \nolinenumbers
%%%%%%%%%%%%%%%%%%%%%%%%%%%%%%%%%%%%%%%%%%%%%%%%%%%%%%%%%%%%%%%%%%%%%
% TABLES
%%%%%%%%%%%%%%%%%%%%%%%%%%%%%%%%%%%%%%%%%%%%%%%%%%%%%%%%%%%%%%%%%%%%%
%% Enter tables at the end of the document, before figures.
\begin{table}[t]
	\caption{List of abbreviations}\label{table:1}
	\begin{center}
		\begin{tabular}{ccccrrcrc}
\topline
 Abbreviations & Full form\\
\midline
			MISO & Monsoon Intra-seasonal oscillations \\
			LPS & Low Pressure System \\
			ERE & Extreme rainfall event \\
			LPS-noex & Low Pressure System that do not produce any Extreme rainfall event \\
			LPS-sm & Small size extreme rainfall event producing low pressure system \\
			LPS-med & Medium size extreme rainfall event producing low pressure system \\
			LPS-lg & Large size extreme rainfall event producing low pressure system \\
			SCV & Secondary cyclonic vortex\\
			LLJ & Low level jet\\
\botline
\end{tabular}
\end{center}
\end{table}

%
%\begin{table}[t]
%\caption{\protect Time-averaged vorticity budget terms. The averages are taken along the track of LPSs from the time of their genesis ($t=0$ $hrs$) to Day(+5) ($t=120$ $hrs$).}
%\begin{center}
%\begin{tabular}{ccccrrcrc}
%\hline\hline
%Terms of vorticity budget($10^{-9} s^{-2}$) & $LPS-noex$ & $LPS-sm$ & $LPS-med$ & $LPS-lg$\\
%\hline
% Growth rate $\left( \frac{\partial \zeta}{\partial t}  \right)$ & -0.014 & 0.013 & 0.047 & 0.131 \\
% Horizontal advection $\left(-V\cdot \nabla (\zeta+f) \right )$ & -0.092 & -0.044 & -0.079 & -0.041 \\
% Vorticity divergence $\left(- (\zeta+f) \nabla \cdot V \right)$ & 0.081 & 0.122 & 0.316 & 0.300 \\
% Vertical advection $\left( \omega \frac{\partial \zeta}{\partial p} \right)$ & 0.092 & 0.074 & 0.165 & 0.199 \\
% tilting  $\left(k \cdot \left( \frac{\partial V}{\partial p} \times \nabla \omega   \right) \right)$ & 0.006 & -0.015 & -0.052 & -0.170\\
% residual& -0.101 & -0.124 & -0.303 & -0.157\\
%\hline
%\end{tabular}
%\end{center}
%\label{\protect table:2}
%\end{table}

%%%%%%%%%%%%%%%%%%%%%%%%%%%%%%%%%%%%%%%%%%%%%%%%%%%%%%%%%%%%%%%%%%%%%
% FIGURES
%%%%%%%%%%%%%%%%%%%%%%%%%%%%%%%%%%%%%%%%%%%%%%%%%%%%%%%%%%%%%%%%%%%%%
%% Enter figures at the end of the document, after tables.
%%

\begin{figure}[h]
	\hspace{-1 cm}	
\centerline{\includegraphics[scale=0.7]{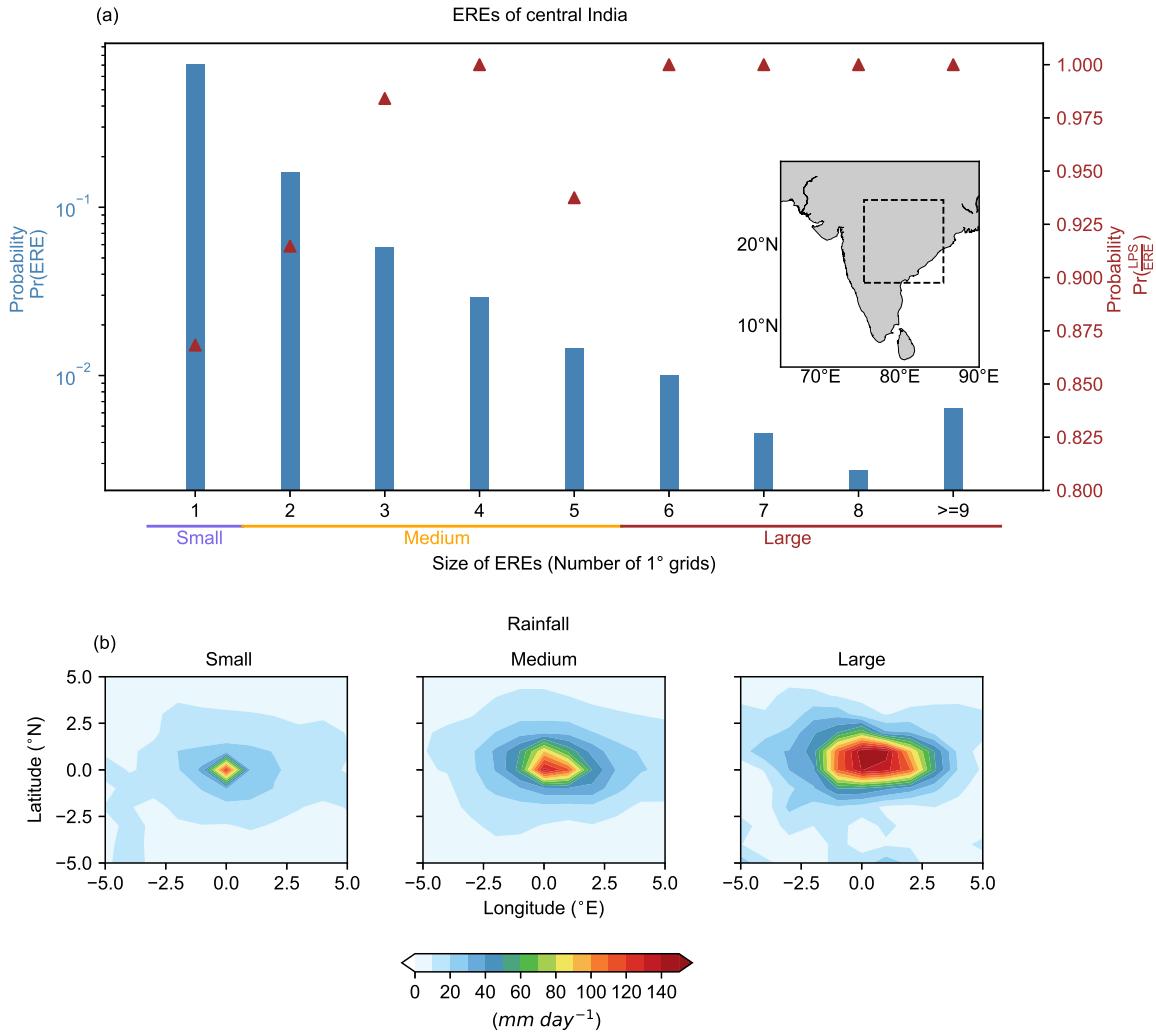}}
    \caption{(a) The probability of occurrence of  Extreme Rainfall Events (EREs) ($Pr(ERE)$, bars), and the probability of a presence of an LPS for the EREs of different sizes ($Pr (\frac{LPS}{ERE})$, brown markers), where the size is measured in $1^{\circ}\times 1^{\circ}$ grid units. Based on the size, the EREs are classified as Small ($size=1$), Medium ($size$ $2-5$) and Large ($size \geq 6$). Note the logarithmic scale for the probability of occurrence. The probabilities for $size\geq 9$ are clubbed in the last bar. (b) Rainfall intensity ($mm$ $day^{-1}$) distribution with respect to the geometric centre of the EREs. The inset box in Fig. 1a shows the study region (Central India: $15^{\circ}$N-$25^{\circ}$N,   $75^{\circ}$E-$85^{\circ}$E).}
\label{f1}		
\end{figure}

\begin{figure}[h]
	\includegraphics[scale=0.8]{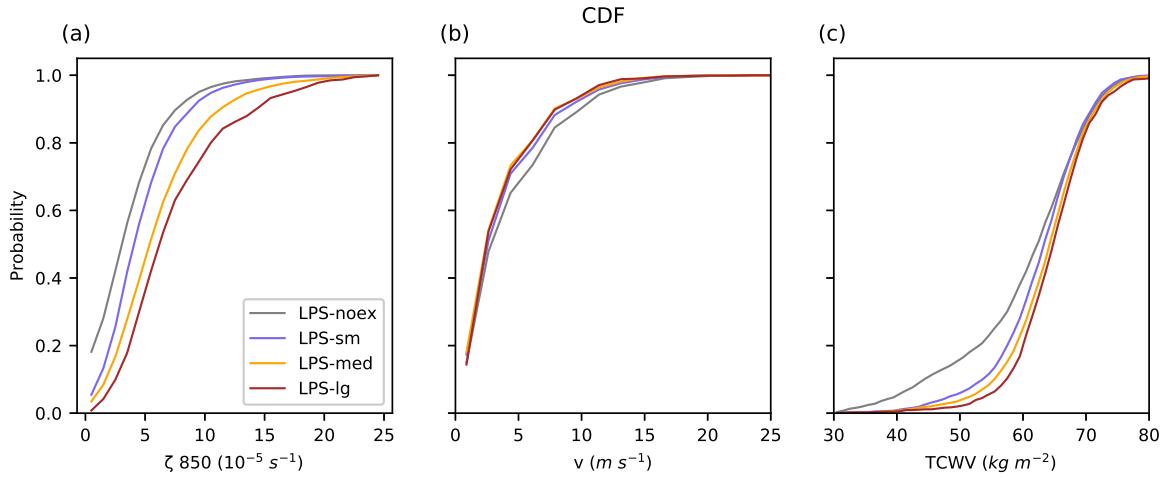}
	\caption{Cumulative distribution function (CDF) of (a) relative vorticity at 850 hPa ($\zeta_{850}$), (b) speed of propagation, and (c) total column water vapor (TCWV) for the LPSs that lead to the EREs of different sizes viz., Small (LPS-sm), Medium (LPS-med) and Large (LPS-lg), and the LPSs that did not lead any ERE over central India (LPS-noex). The 6-hourly values along the tracks of LPSs are used to calculate the CDFs.}\label{f2}	
\end{figure}

\begin{figure}[h]
	\includegraphics[scale=0.8]{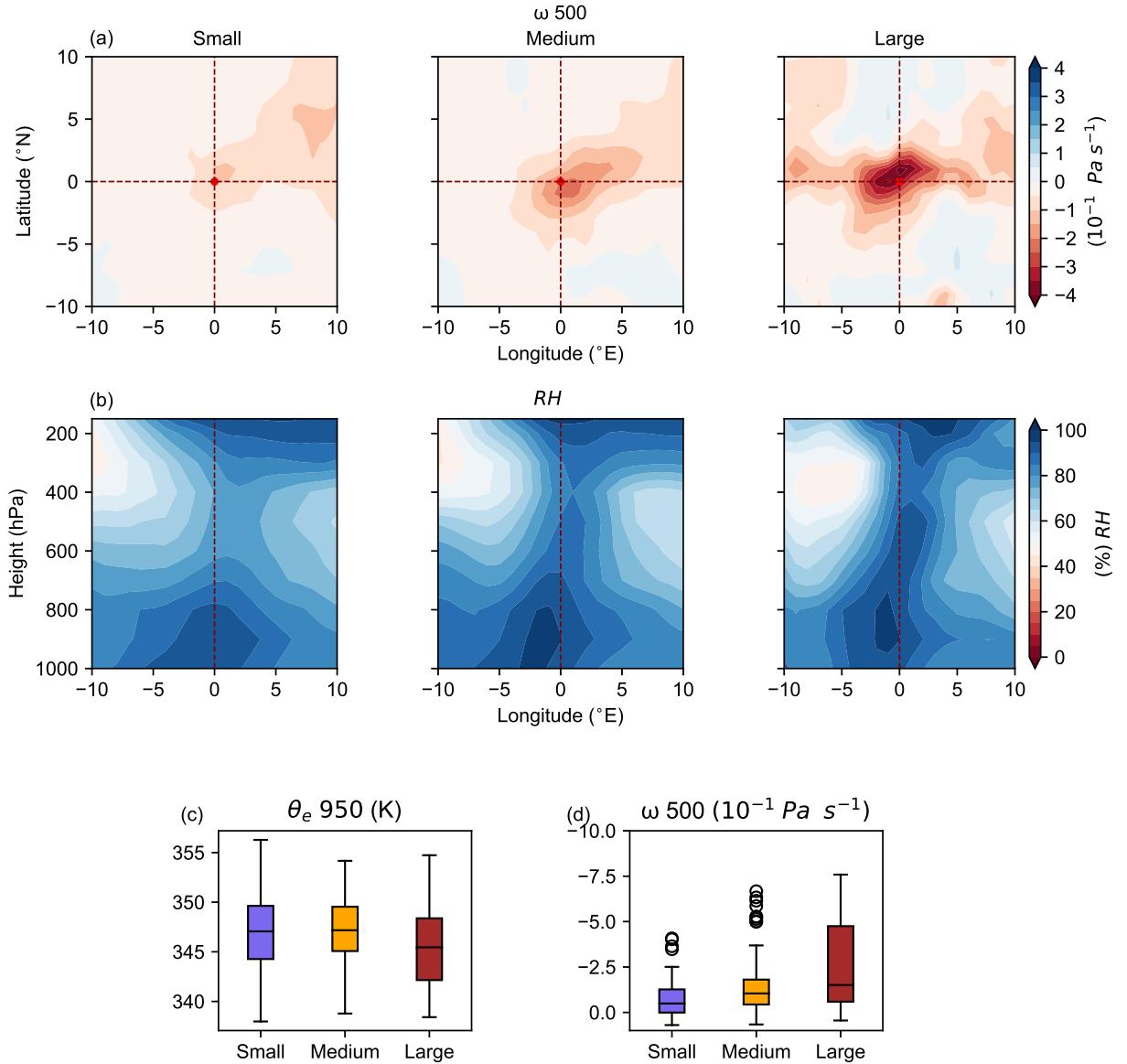}
	\vspace{-1.5 cm}
	\caption{Distribution of the daily composite (a) pressure velocity at 500 hPa ($\omega 500$) with respect to the geometric centre of the EREs, (b) Relative Humidity ($RH$) in the longitude-height plane along the latitude of the EREs (The red dashed horizontal line in Fig 3a). The red dashed horizontal and vertical lines pass through the centre of the ERE. Box plot of the average (c) equivalent potential temperature at 950 hPa ($\theta_{e} 950$) and (d) $\omega$ at 500 hPa ($\omega 500$ ). The average of $\theta_{e} 950$ and $\omega 500$ are calculated over $2^{\circ}\times 2^{\circ}$ latitude-longitude box around the geometric centre of EREs. The horizontal black lines in the box plot indicate the median values. The box shows the inter-quartile range. The whiskers and black circles denote the extreme values (non-outliers) and outliers, respectively.}\label{f3}		
	
\end{figure}

\begin{figure}[h]
	\hspace{-1.5 cm}	
	\includegraphics[scale=0.7]{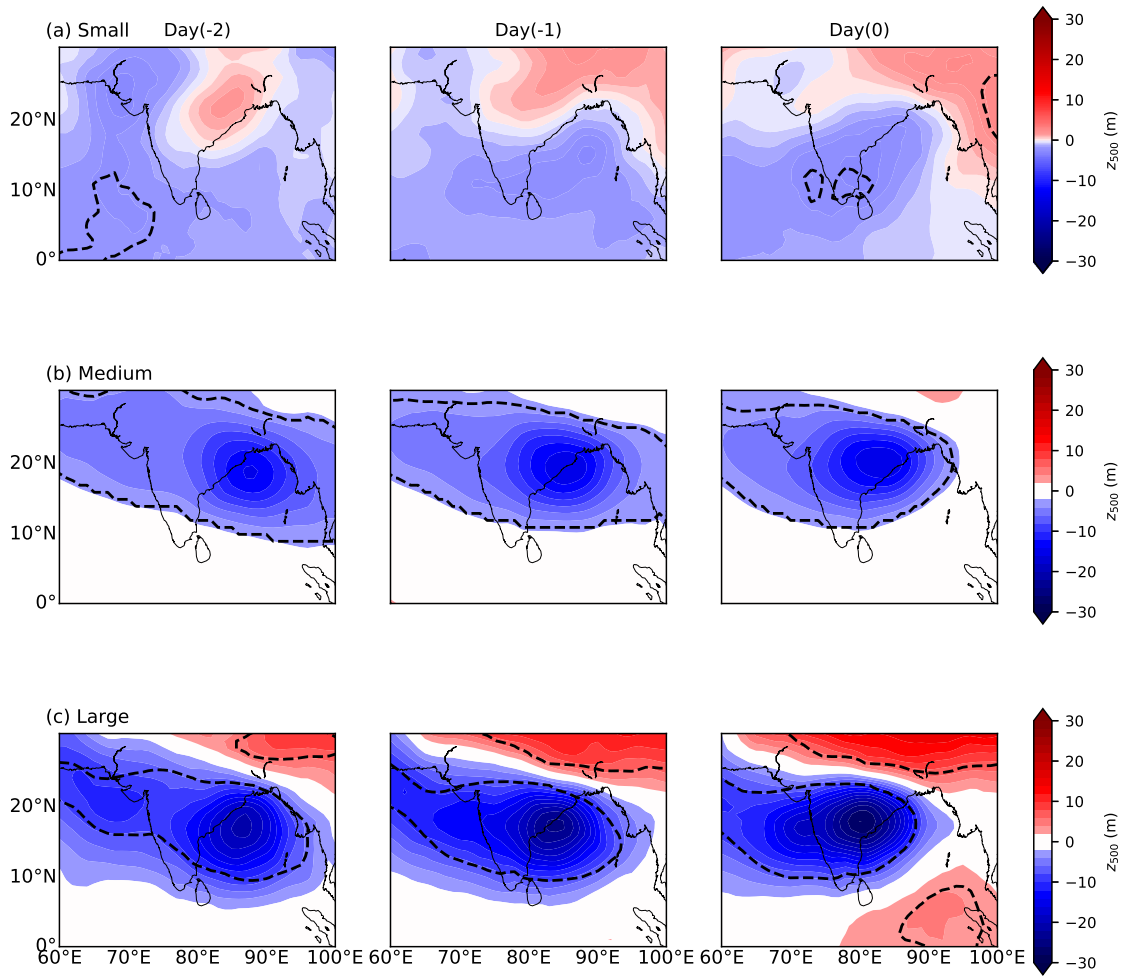}
	\caption{Lagged composites of anomalous geopotential height ($m$) at 500 hPa for (a) Small (b) Medium, and (c) Large EREs from Day(-2) to Day(0) (left to right). Day(0) indicates the day of an ERE. The dashed contour represent statistically significant anomalies of geopotential height ($t-test: pvalue < 0.05$).}
	\label{f4}		
\end{figure}

\begin{figure}
	\includegraphics[scale=0.8]{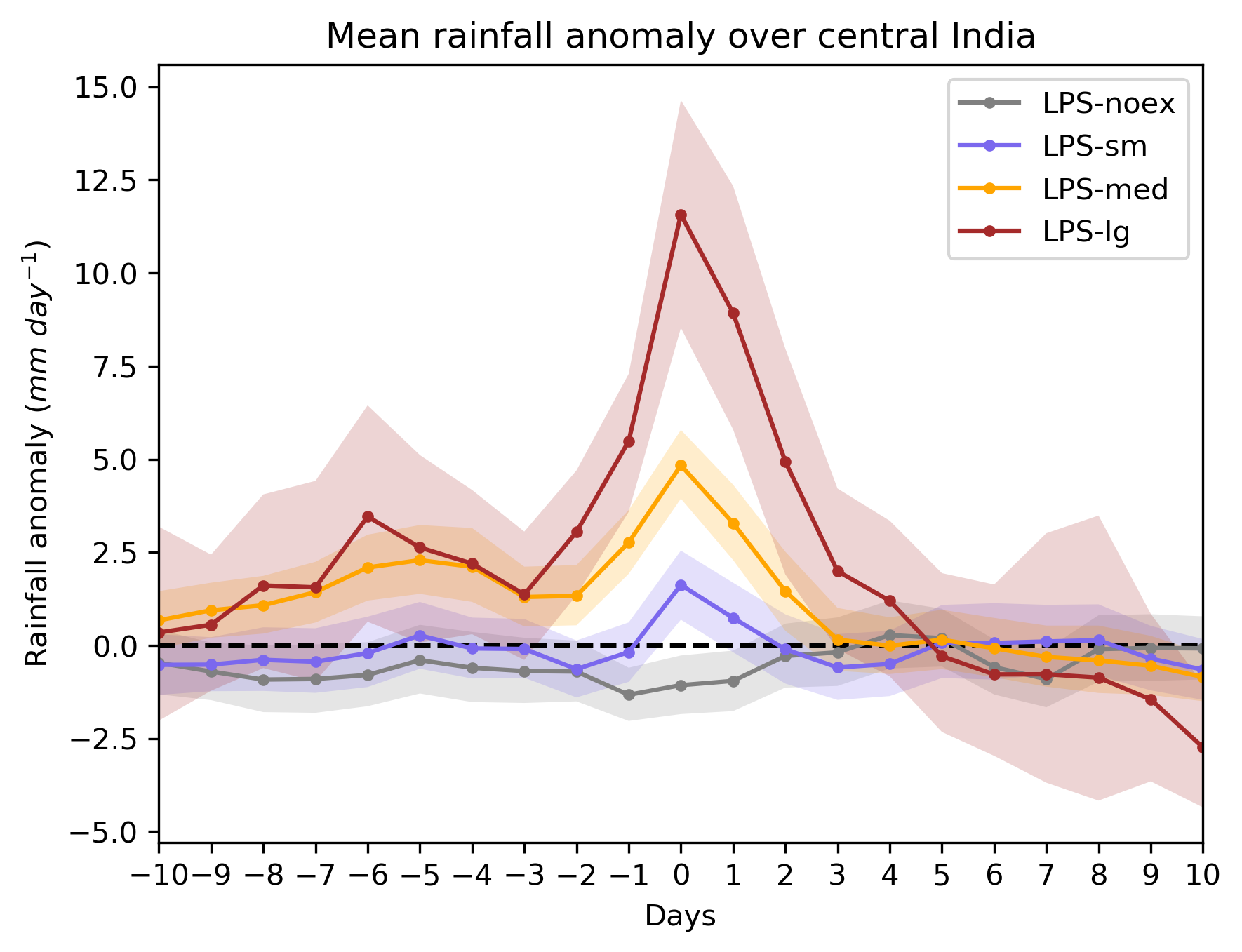}
	\caption{Composite daily rainfall anomalies over the core monsoon zone ($15^{\circ}$-$25^{\circ}$N, $65^{\circ}$-$75^{\circ}$E) for lag days -10 to +10 for LPS-sm, LPS-med, LPS-lg, and LPS-noex. The shading indicates the 95\% confidence interval obtained by generating $10^{4}$ bootstrapping samples. The black dashed line indicates zero line.}
	\label{fig:fig5}		
\end{figure}

\begin{figure}	
	\includegraphics[scale=0.8]{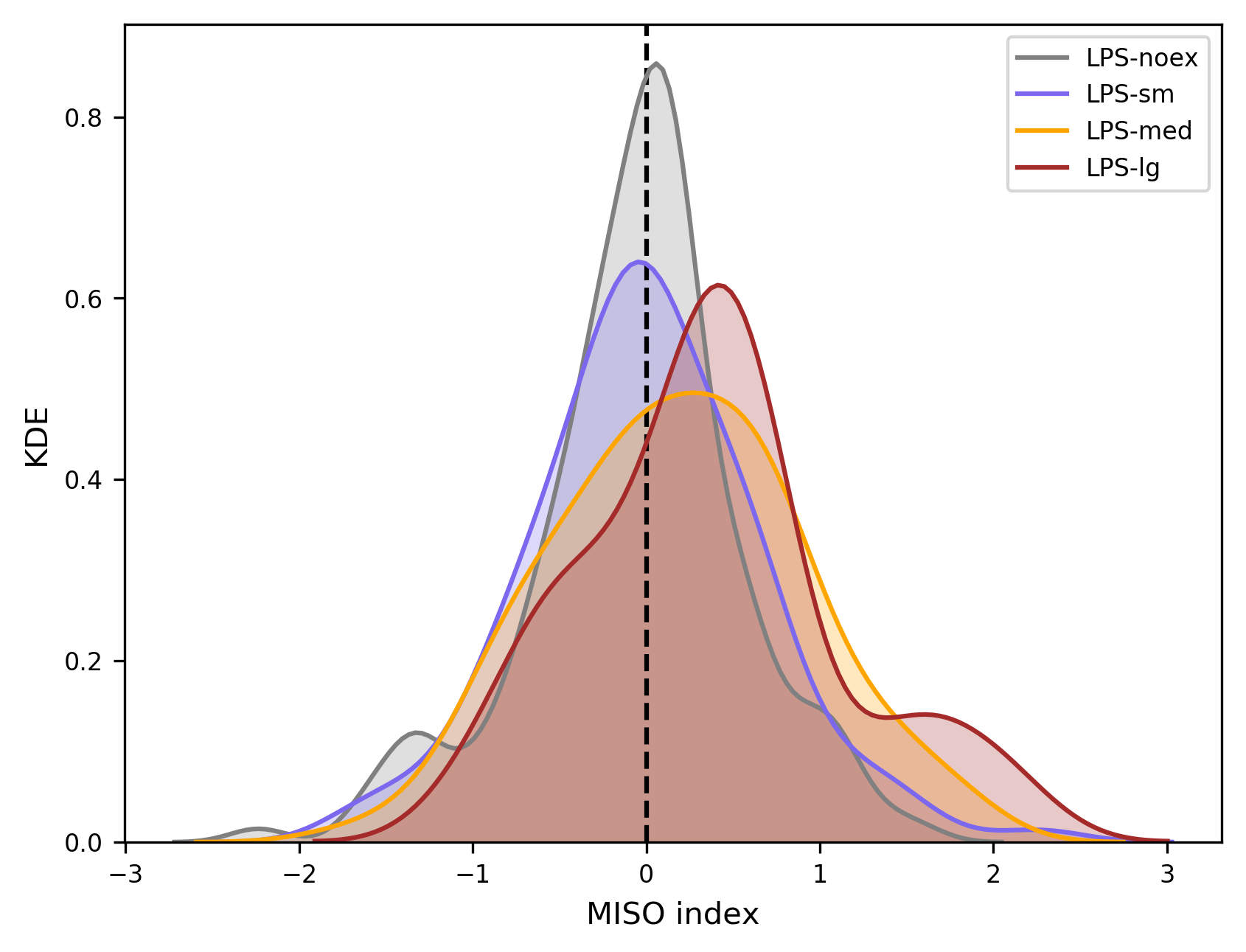}
	\caption{Kernel density estimation (KDE) of the MISO index (defined in section 3b) on Day(0) of LPS-sm, LPS-med, LPS-lg and LPS-noex.}
	\label{fig:fig6}		
\end{figure}

\begin{figure}
	\includegraphics[scale=0.8]{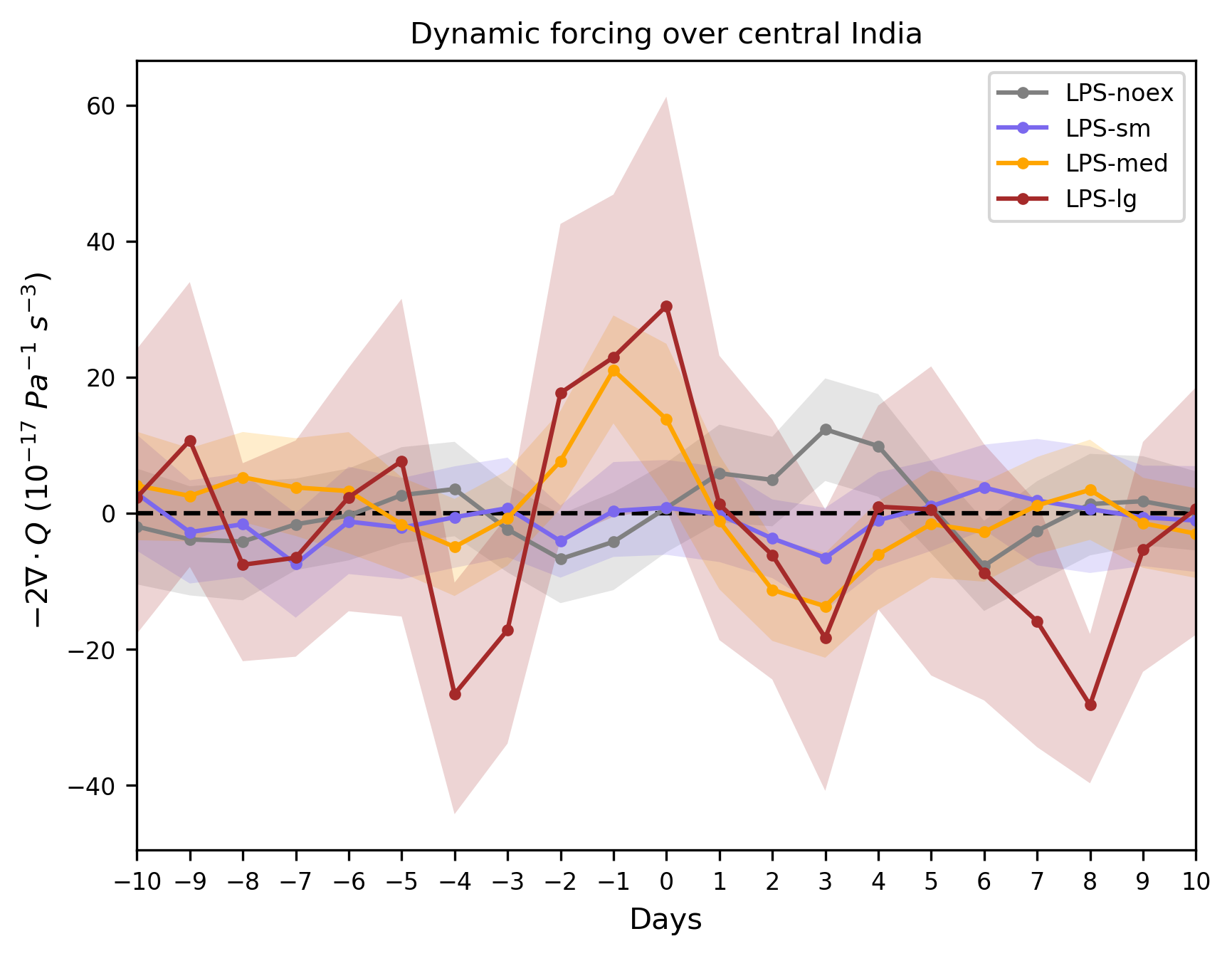}
	\caption{Composite daily QG forcing ($-2 \nabla \cdot Q$) over the study region for lag days -10 to +10. The shading indicates the 95\% confidence interval obtained by generating $10^{4}$ bootstrapping samples.}
	\label{fig:fig7}		
\end{figure}

\begin{figure}
%	\hspace{-1.5 cm}	
	\includegraphics[scale=0.7]{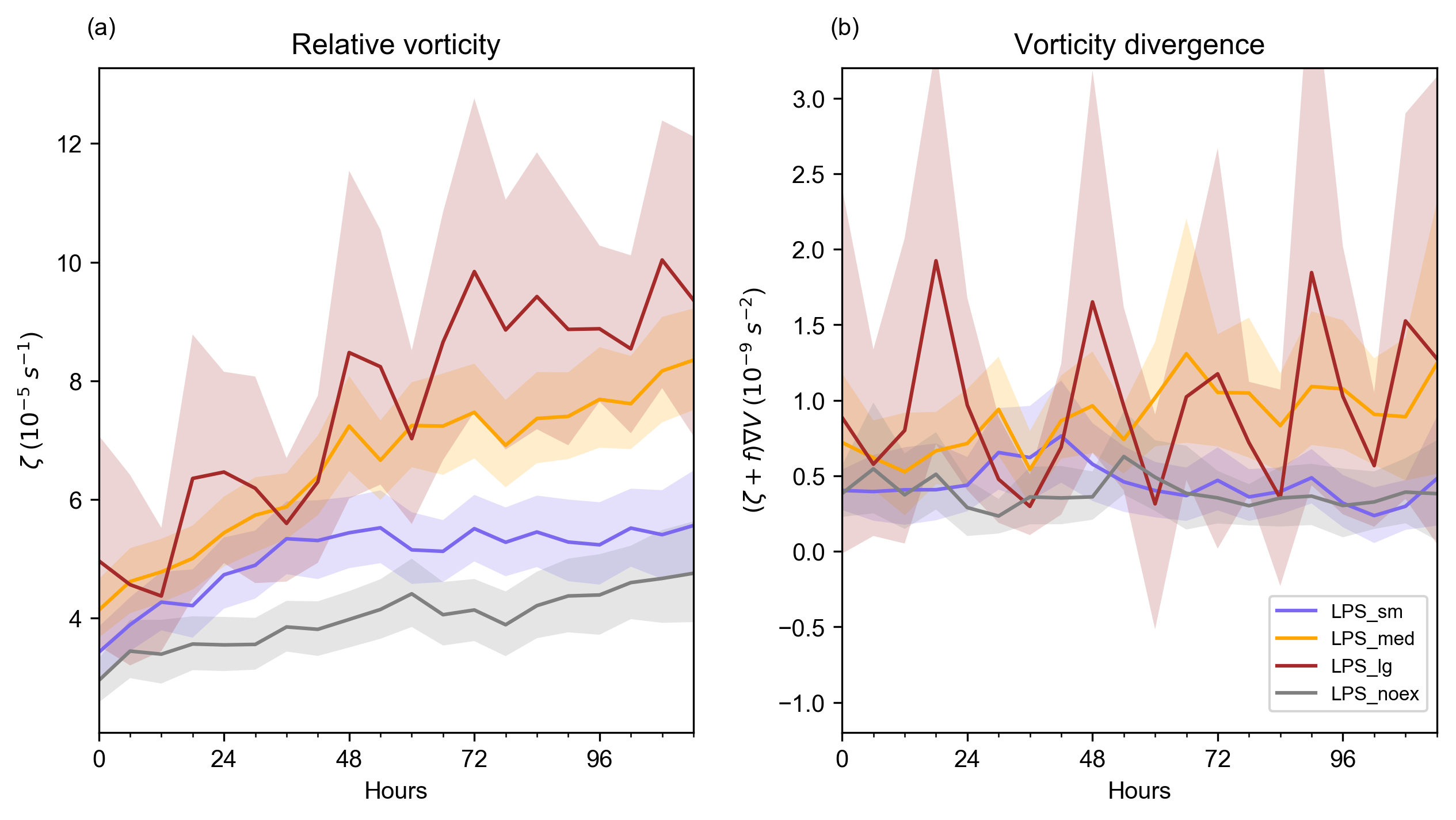}
	\caption{Time evolution of (a) relative vorticity and (b) the vorticity divergence at 850 hpa. The 6-hourly terms are evaluated along the tracks of LPSs. The shading indicates the 95\% confidence interval obtained by generating $10^{4}$ bootstrapping samples.}
	\label{fig:fig8}		
\end{figure}

\begin{figure}
		\hspace{-1.5 cm}	
	\includegraphics[scale=0.6]{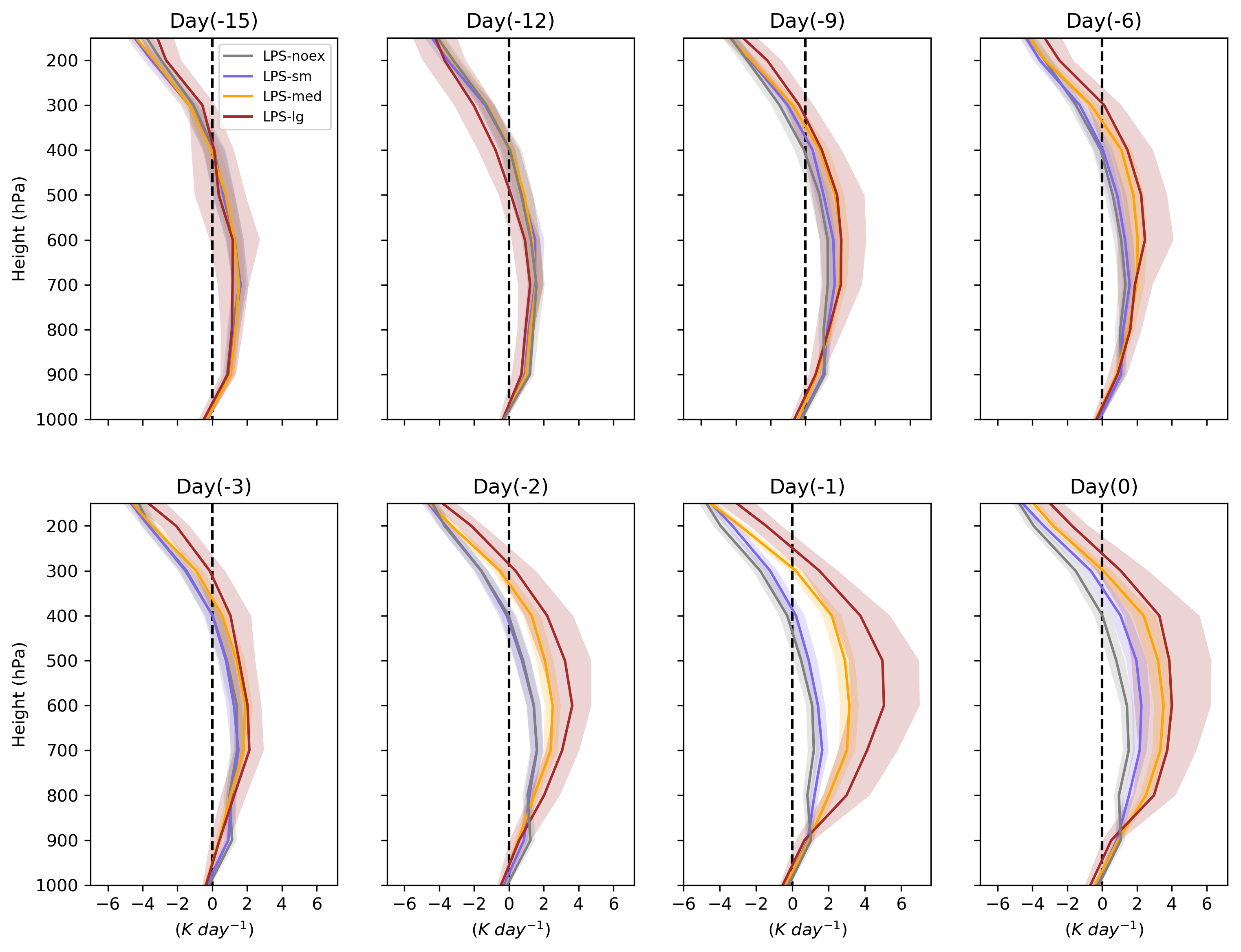}
	\caption{Vertical profile of diabatic heating averaged over the study region from Day(-15) to Day(0). The shading represents 95\% confidence interval obtained by generating $10^{4}$ bootstrapping samples.}
	\label{fig:fig9}		
\end{figure}

\begin{figure}
	\hspace{-1.5 cm}	
	\includegraphics[scale=0.6]{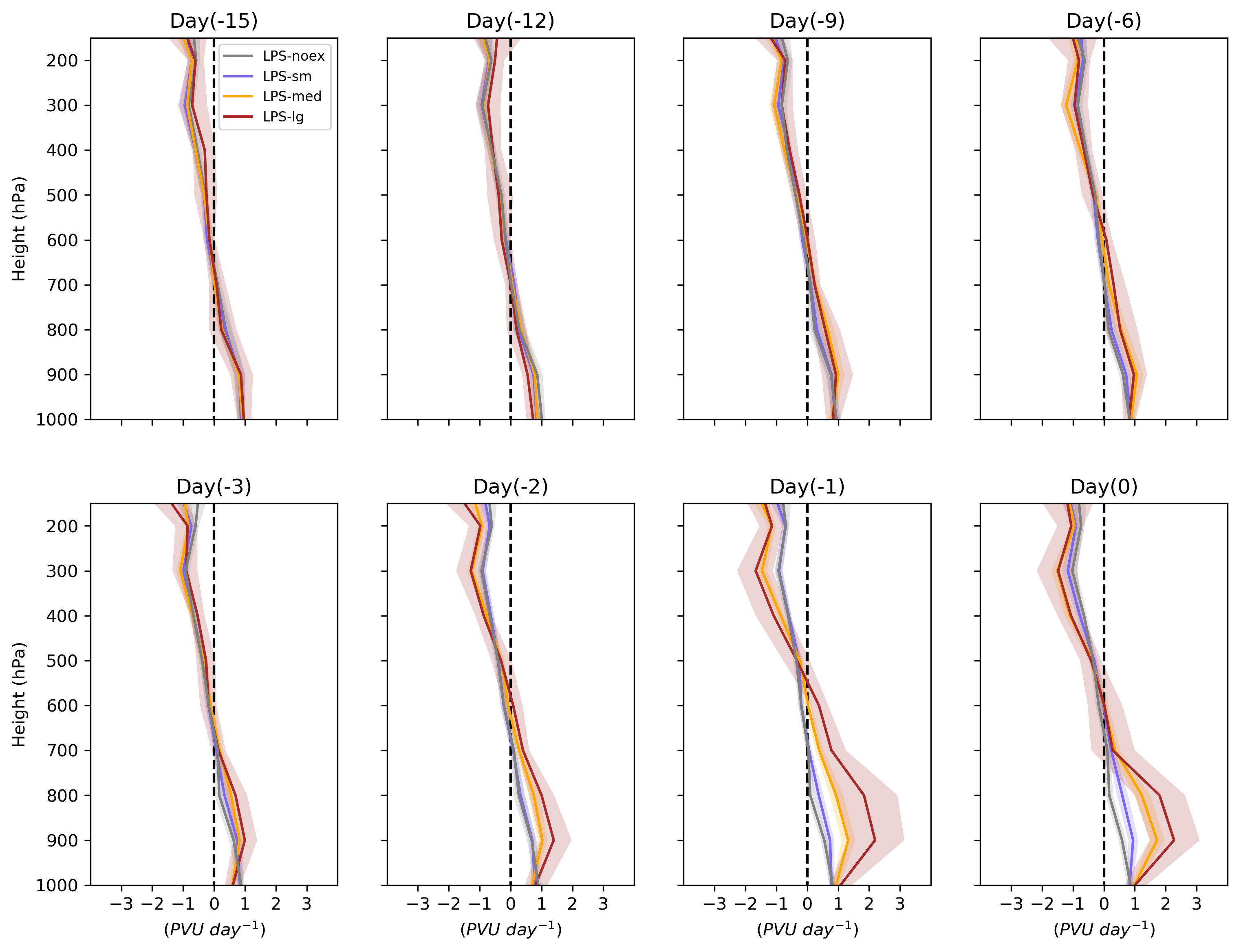}
	\caption{ Same as Fig. 9, except for the diabatic PV tendency (obtained by equation(3)).}
	\label{fig:fig10}		
\end{figure}

%\begin{figure}
%	\hspace{-1.6 cm}	
%	\includegraphics[scale=0.4]{Fig12summary.png}
%	\caption{The flowchart explaining the processes through which the phases of the MISO interact with monsoon low pressure systems and produce EREs of different spatial extents.}
%	\label{fig:fig11}		
%\end{figure}

\end{document}